\documentclass{PoS}

\title{Ten years of the INTEGRAL Burst Alert System (IBAS)}

\ShortTitle{Ten years of IBAS}

\author{\speaker{Sandro Mereghetti}\\
        IASF-Milano, INAF\\
        via E.~Bassini, 15, I 20133 Milano, Italy\\
        E-mail: \email{sandro@iasf-milano.inaf.it}}

\abstract{The INTEGRAL Burst Alert System (IBAS) has been developed to detect and locate in real time the  gamma-ray bursts (GRBs)
serendipitously observed by INTEGRAL. 
The IBAS  software runs automatically at the INTEGRAL Science Data Centre (ISDC), where the satellite data are received with a delay of only a few seconds.  
The sky coordinates of the GRBs occurring in the field of view of  the IBIS instrument are distributed via Internet in real time. 
The localizations have a typical uncertainty radius of 2 arcmin (90\% c.l.) and in most cases are available within a few tens of seconds after the beginning of the GRB. 
In  ten years of operations IBAS has localized about 90 GRBs, most of which in near real time, and distributed alerts also for other kinds
of astrophysical transient events, such as  type I bursts from low mass X-ray binaries, flares and bursts from magnetars, and outbursts of Galactic transients.
IBAS also provides the light curves for the GRBs detected with the anti-coincidence shield of the SPI instrument.
Here I summarize the main properties of the GRBs detected in the field of view of IBIS during the first ten years of
the INTEGRAL mission.
}

\FullConference{"An INTEGRAL view of the high-energy sky (the first 10 years)"
9th INTEGRAL Workshop and celebration of the 10th anniversary of the launch,\\
		October 15-19, 2012\\
		Bibliotheque Nationale de France, Paris, France}

\def\approxgt{\mathrel{\hbox{\rlap{\lower.55ex \hbox {$\sim$}}
        \kern-.3em \raise.4ex \hbox{$>$}}}}
\def\approxlt{\mathrel{\hbox{\rlap{\lower.55ex \hbox {$\sim$}}
        \kern-.3em \raise.4ex \hbox{$<$}}}}

\def\ltsima{$\; \buildrel < \over \sim \;$}
\def\lsim{\lower.5ex\hbox{\ltsima}}
\def\gtsima{$\; \buildrel > \over \sim \;$}
\def\gsim{\lower.5ex\hbox{\gtsima}}

\begin{document}

\section{Introduction}
 
The IBAS  software  \cite{mer03} runs in real time at the INTEGRAL Science Data Centre \cite{cou03}, where the
INTEGRAL  telemetry is continuously received from the Mission Operation Center on a 128 kbs dedicated line.
IBAS is based on a flexible multi-thread architecture that allows different triggering algorithms to 
operate in parallel on the IBIS/ISGRI data in order to detect, localize and validate GRBs (and other transients)
in the field of view of IBIS. Alerts with the positions of GRBs detected with high significance are generated
and distributed automatically. Quick-look  interactive analysis to confirm the triggers and derive 
refined GRB parameters is carried out at IASF-Milano and CEA-Saclay. Lower significance triggers
are also checked and may in some case lead to manually generated alerts.
Thanks to IBAS, INTEGRAL  has been the first mission to distribute in real time the positions of GRBs with arcminute  accuracy. 

Since  the beginning of 2011, alerts for  triggers
of intermediate significance  are also distributed automatically to the IBAS users who  request them, 
e.g. for use  with robotic telescopes.  Although  most of these triggers  cannot be confirmed as 
GRBs, based only on INTEGRAL data, they might be validated by the presence of optical and/or X-ray counterparts.
Note that two of the four short GRBs detected by INTEGRAL  (GRB 100703A and 110112B) produced
triggers of intermediate significance, that were subsequently confirmed by an interactive analysis.

In the following I summarize  some properties of all the GRBs detected up to now (2012 November) in the field of view of the IBIS instrument.
More details on the global properties of a smaller sample ($\sim$2002-2007) can be found in \cite{via09} and \cite{fol08}.
For updated information on IBAS and on GRBs detected by INTEGRAL see also \textit{http://ibas.iasf-milano.inaf.it/}.

 \begin{figure}
\hbox{
\includegraphics[width=.5\textwidth]{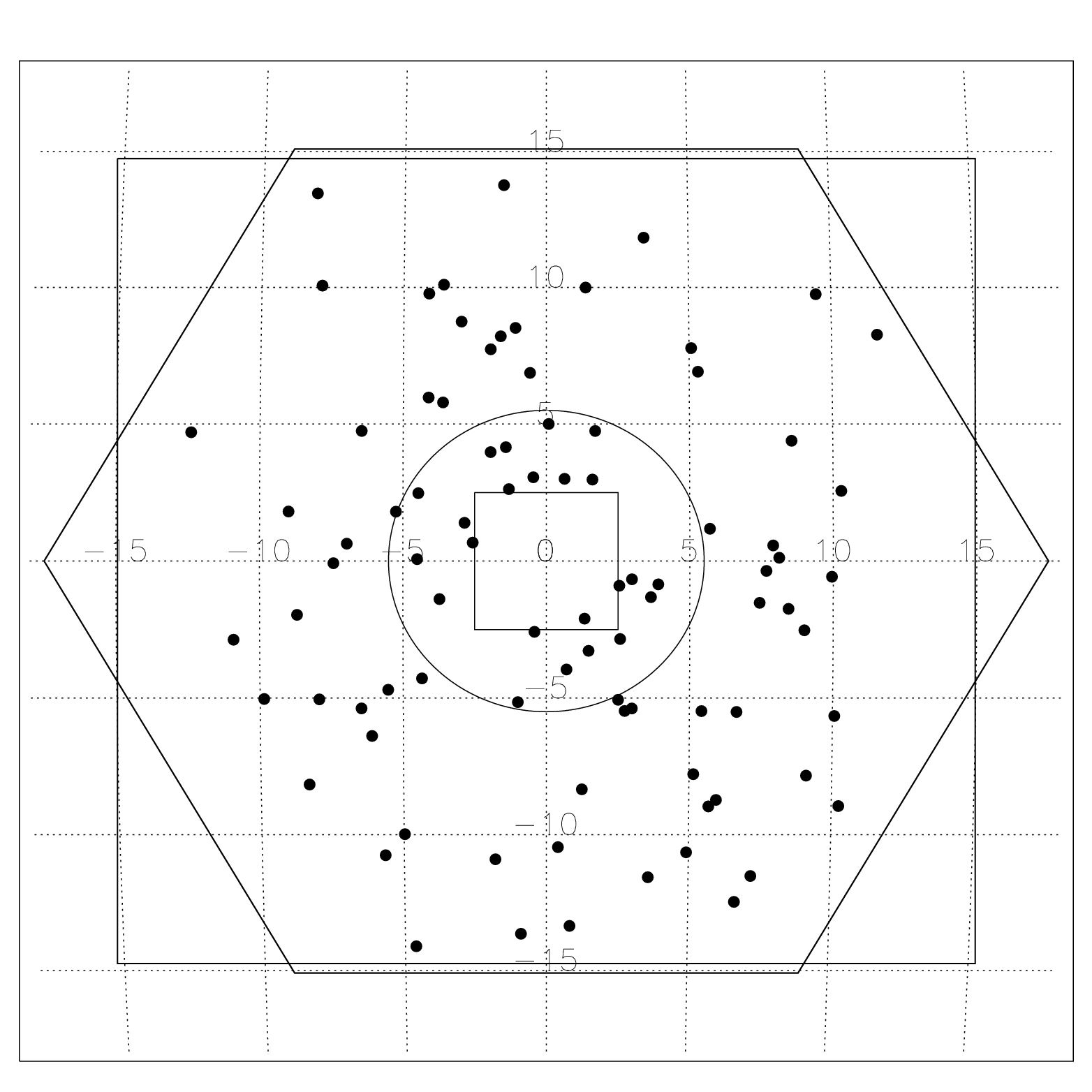}
\includegraphics[width=.5\textwidth]{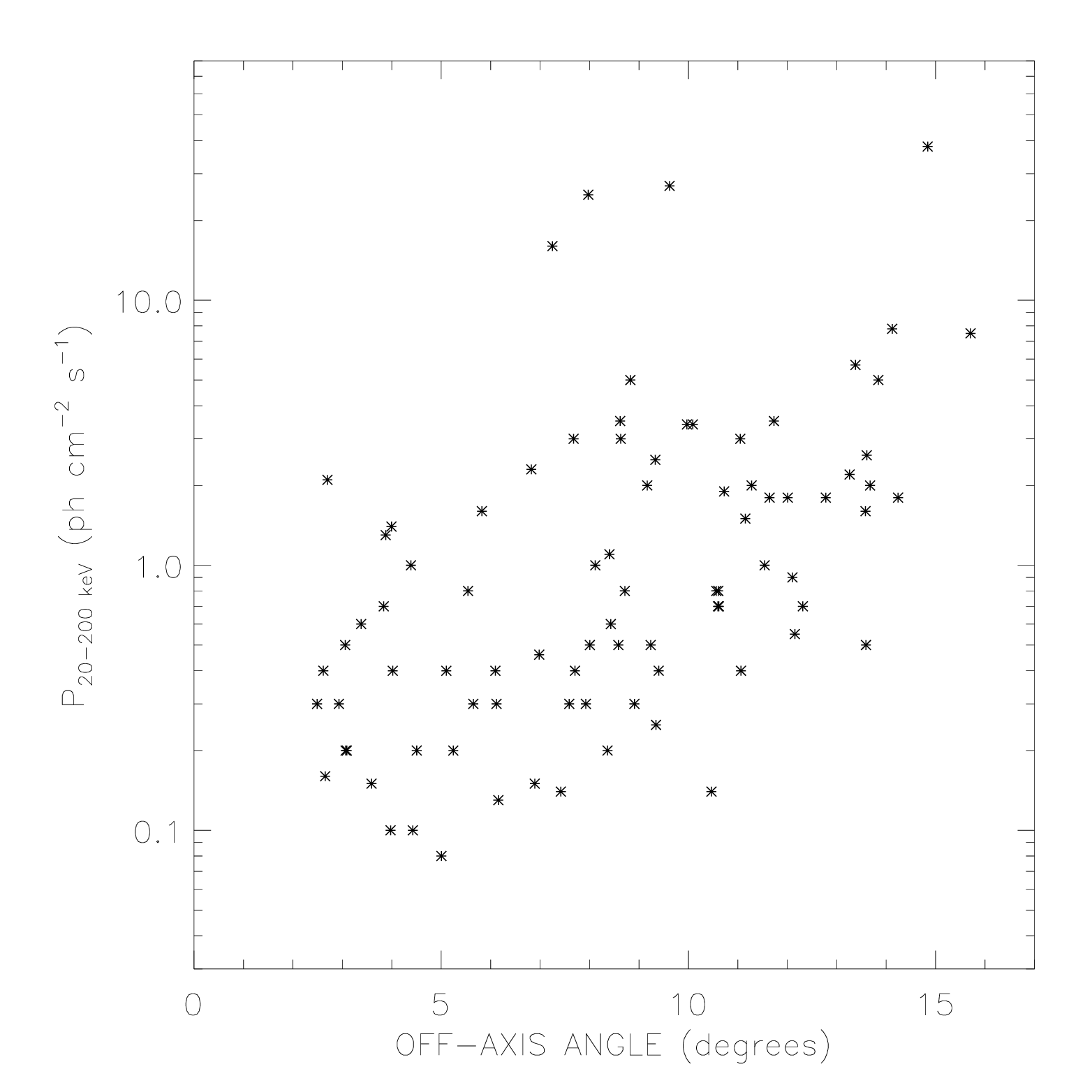}}
\caption{\textit{Left panel:} Positions of the GRBs in the fields of view of IBIS \cite{ibis} (large square), SPI \cite{spi} (hexagon), JEM-X \cite{jemx} (circle) and OMC \cite{omc} (small square).
\textit{Right panel:} GRB peak flux (20-200 keV) as a function of the off-axis angle. }
\label{fig1}
\end{figure}

\begin{figure}
\hbox{
\includegraphics[width=.5\textwidth]{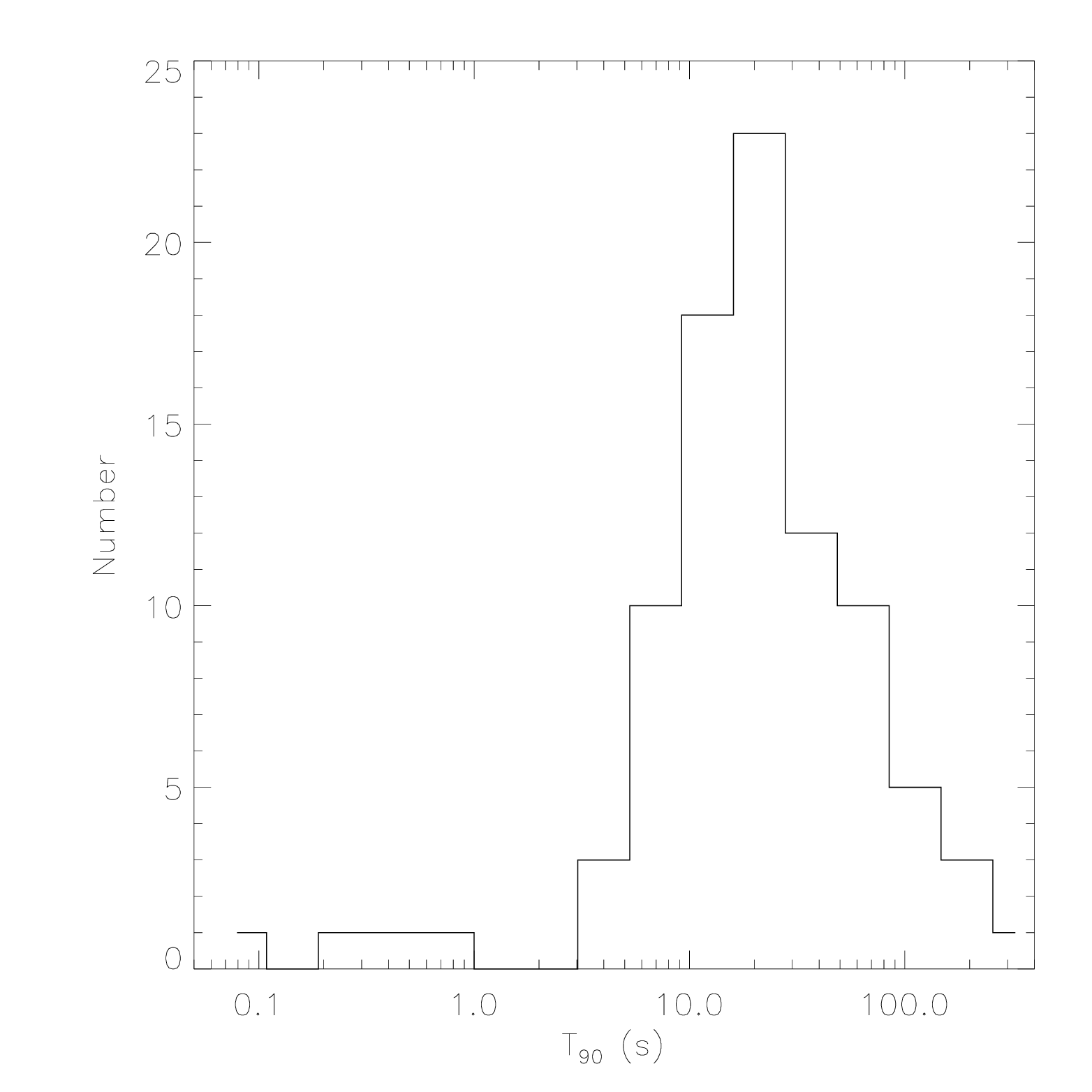} 
\includegraphics[width=.5\textwidth]{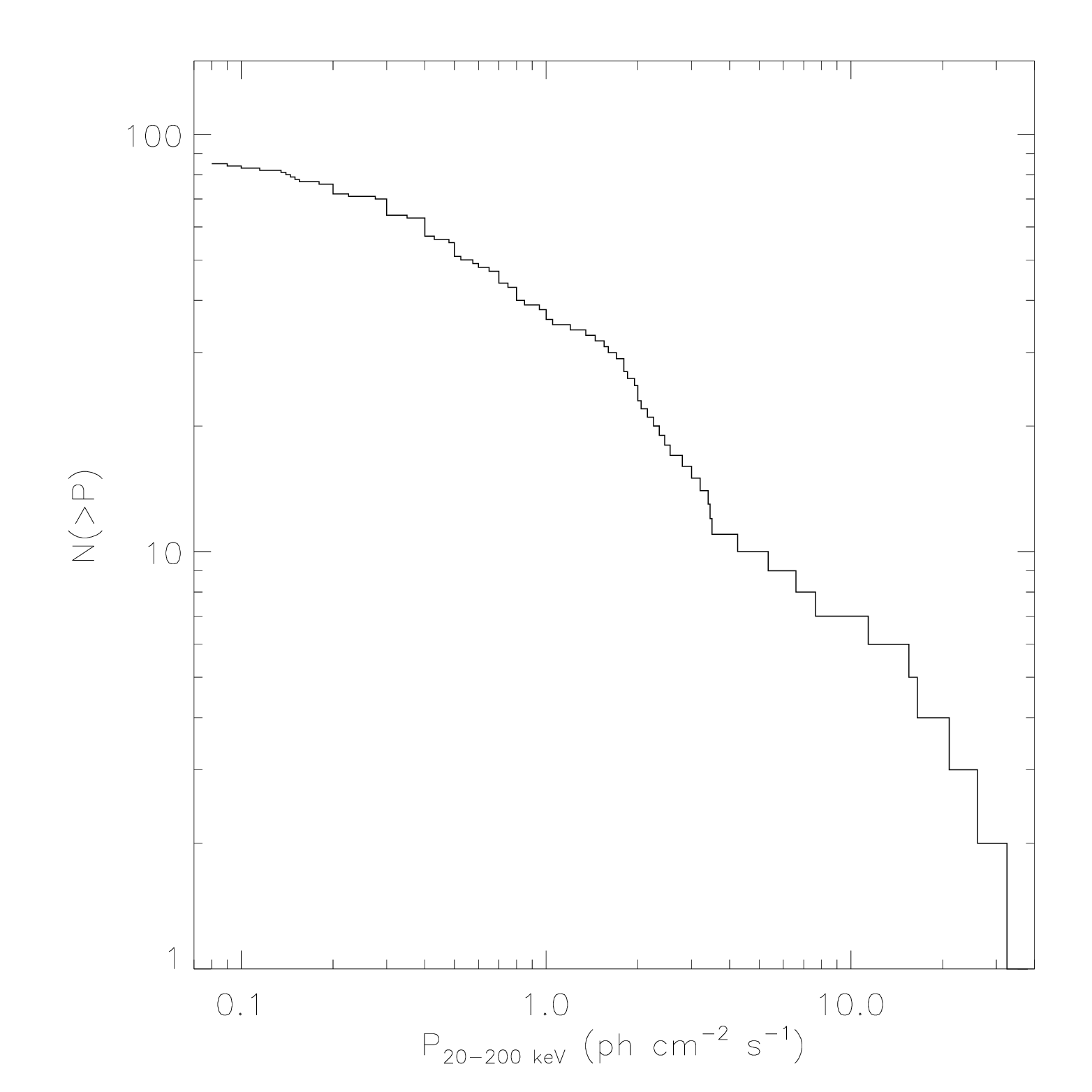}}
\caption{\textit{Left panel:} Distribution of the GRB durations.
\textit{Right panel:} LogN-LogP integral distribution of the long GRBs detected with IBIS. }
\label{fig2}
\end{figure}

\section{GRBs in the IBIS field of view}

During the first ten years of the INTEGRAL mission, 89 GRBs have been detected in the field of view
of IBIS. Some of their properties are listed in  Table 1.
The Table gives  the approximate start time, T$_{90}$ duration, and peak flux in the 20-200 keV energy range for each burst.
The peak fluxes refer to an integration time of 1 s, except for the four short bursts (as indicated in the \textit{Notes} column).
The detection of a radio (R), optical (O) or X-ray (X) afterglow is   indicated  in the \textit{Cpts} column (lower case letters
refer to non-confirmed counterparts). The  \textit{IBAS} column indicates whether the GRB position was automatically distributed
in real time (RT) or with a delay (D) after interactive verification.
Two events initially reported as GRBs are not included in the table:  they are GRB 071017, which is positionally consistent with the
X-ray source AX J1818.8--1559 and is most likely a Galactic soft gamma-ray 
repeater \cite{mer12}, and GRB 120118A, probably a spurious trigger caused by the X-ray binary GX 301--4.

Figure 1  shows the positions of the GRBs in the field of view of the INTEGRAL instruments  (left panel)
and their peak fluxes as a function of the off-axis angle  (right panel). The triggering sensitivity is practically
constant within the IBIS fully coded field of view (the central $\sim9^{\circ}\times9^{\circ}$), while it decreases at larger
off-axis angles. Note that the size of the error region  depends only on the statistical significance of the source detection,
therefore  accurate localizations  can  be obtained also for the  GRBs detected at very large off-axis angles.
Only one GRB occurred inside the small field of view of the OMC (Optical Monitoring Camera) instrument ($\sim5^{\circ}\times5^{\circ}$),
but unfortunately it was located very close to one of the brightest stars in the sky, which caused
saturation of the OMC image at the burst position. 

The distribution of GRB durations is shown in the left panel of Fig.2. IBIS detected four short bursts: GRB 070707, GRB 081226B, GRB 100703A, and GRB 110112B.

The right panel of Fig.2  shows the integral distribution of the peak fluxes of the 85 long GRBs. A maximum likelihood estimate \cite{cra70},
excluding the region below 0.4 photons cm$^{-2}$ s$^{-2}$ which is affected by incompleteness,  indicates that the distribution
can be described by a power-law  N($>$P) $\propto$ P$^{-\alpha}$  with $\alpha$=0.6$\pm$0.1.

\begin{table}
\begin{footnotesize}
\begin{tabular}{llrclll}
\hline
    GRB & Time     &T$_{90}$ &  Peak Flux           & Ctpts & IBAS & Notes             \\
        &  UT      &    s    &ph cm$^{-2}$ s$^{-1}$ &       &      &                   \\
\hline
021125  & 17:58:25 &   24   &    16.              &  -  &  D  &  During IBIS Performance/Verification Phase\\   
021219  & 07:33:54 &   5    &    3.4              &  -  &  D  & \\ 
030131  & 07:38:49 &  124   &   1.9               &  O  &  D  & Detected during a satellite slew \\ 
030227  & 08:42:02 &   15   &   0.8               &  OX  &  D  & Alerts disabled during Crab observ.  \\ 
030320  & 10:11:49 &  48    &   7.5               &  -  &  D  & Below   threshold for automatic alert delivery.\\ 
030501  & 03:10:02 &  25    &    2.2              &  -  &  RT & \\ 
030529  & 19:53:18 &  16    &   0.14              &  -  &  D  &  during solar flare; detected off-line  \\
031203  & 22:01:27 &  19    &   1.5               &  ROX  &  RT & z=0.105, SN 2003lw   \\
040106  & 17:55:10 &  48    &   0.7               &  OX  &  RT & \\
040223  & 13:24:51 & 198    &   0.3               &  X  &  RT & \\
040323  & 13:02:58 &  14    &   1.8               &   o  &  RT & \\ 
040403  & 05:08:03 &  15    &   0.4               &  -  &  RT & \\
040422  & 06:57:59 &  4     &   2.5               &  O  &  RT & \\
040624  & 08:21:35 & 27     &   0.55              &  -  &  D  & Below   threshold for automatic alert delivery \\
040730  & 02:12:06 & 42     &   0.3               &  -  &  RT & \\
040812  & 06:01:52 &  8     &   0.6               &   oX &  RT & \\
040827  & 11:50:50 & 32     &   0.7               &  OX  &  D  & Below   threshold for automatic alert delivery\\
040903  & 18:17:58 &  7     &   0.3               &  -  &  RT & \\
041015  & 11:11:33 & 30     &   0.25              &  -  &  D  &  Below   threshold for automatic alert delivery \\
041218  & 15:45:44 & 38     &   2.6               &  O  &  RT & \\
041219A & 01:42:13 & 239     &  $>$15              &  RO  &  RT & T$_{90}$=460 including precursor \\
050129  & 20:03:05 & 30     &   0.3               & -   &  D  & Below   threshold for automatic alert delivery \\
050223  & 03:09:00 & 30     &   0.5               & X   &  D  & z=0.584? - one IBIS module off \\
050502  & 02:14:00 & $>$11   &   1.4               &  O  &  RT & z = 3.793? - Started during satellite slew \\
050504  & 08:00:50 & 44     &   0.4               &  X  &  RT & \\
050520  & 00:05:57 & 52     &   1.0               &  X  &  RT & \\
050522  & 06:00:21 & 11     &   0.2               &  x &  RT & \\
050525A & 00:02:53 & 9      &  38.                &  OX  &  D  &  z=0.6; detected off-line (very off-axis)\\
050626  & 03:46:07 & 56     &   0.3               &  -  &  RT & In OMC field of view,  at 2$'$   from $\alpha$ Crucis\\
050714  & 00:05:53 &  34    &   0.4               &  oX &  RT & \\
050918  & 15:36:38 &  280   &   1.8               &  x &  D  & Below   threshold for automatic alert delivery.\\
050922  & 13:43:20 & 10     &   0.1               &  -  &  D  & Below   threshold for automatic alert delivery.\\
051105B & 11:05:41 &   14   &   0.4               &  -  &  RT & \\
051211B & 22:06:07 &   47   &    0.8              &  OX  &  RT & \\
060114  & 12:39:31 &  80    &   0.16              &  -  &  RT & \\
060130  & 04:56:29 &   19   &    0.2              &  -  &  D  & Below   threshold for automatic alert delivery.\\
060204A & 13:19:39 &  52    &   0.2               &  -  &  RT & \\
060428C & 02:30:35 &  10    &   3.5               &  -  &  -  & Detected off-line - IBAS temporarily not running  \\
060901  & 18:43:55 &  16    &    7.8              &  oX &  RT & \\
060912B & 17:32:11 &  140   &   0.15              &  -  &  RT & \\
060930  & 09:04:12 &   9    &   0.6               &  -  &  RT & \\
061025  & 18:35:53 & 11     &    1.1              &  OX  &  RT & \\
061122  & 07:56:50 &  12    &   $>$17             &  OX  &  RT & \\
\hline
 \end{tabular}
\end{footnotesize}
\caption{GRBs in the IBIS field of view.}
\label{tab1a}
\end{table}

\begin{table}
\begin{footnotesize}
\begin{tabular}{llrclll}
\hline
 GRB    & Time      &T$_{90}$&  PF                  & Ctpts& IBAS& Notes             \\
        &  UT       &  s     &ph cm$^{-2}$ s$^{-1}$ &      &     &                   \\
\hline
070309  & 10:00:39 &  22    &    0.3              &  X  & RT  & \\
070311  & 01:52:34 &  32    &    0.9              &  OX  & RT  & \\
070615  & 02:20:37 &  15    &    0.5              &  X  & RT  & \\
070707  & 16:08:38 &  0.7    &    1.8              & OX   & RT  & Short (peak flux on 0.7 s) \\
070925  & 15:52:32 &   19   & $>$2                & X   & RT  & \\
071003  & 07:40:55 & 148    &     5.7             & OX   &  D  &  One IBIS module off  \\
071109  & 20:35:55 &  30    & 0.46                &  r & RT  & \\
080120  & 17:28:28 &  15    &   3.                &  OX  & RT  & \\
080414  & 22:33:30 &   8    &     1.              &  -  & RT  & \\
080603A & 11:18:15 &  150   &  0.5                &  OX  & RT  & z=1.688\\ 
080613A & 09:35:21 &  30    &    1.3              &  OX  & RT  & \\ 
080723B & 13:22:15 &  95    &  25.                &  X  & RT  & \\ 
080922  & 11:03:36 &   60   &  1.                 &  -  & RT  & \\ 
081003A & 13:46:00 &   15   &   0.3               &  X  & RT  & \\ 
081003B & 20:48:08 &   20   & 3.                  &  -  & RT  & \\
081016  & 06:51:31 &  30    &   $>$2.4            &  X  & RT  & \\ 
081204  & 16:44:55 &  12    &    0.7              &  X  & RT  & \\ 
081226B & 12:13:11 &  0.5    &     3.              &  -  & RT  & Short  (peak flux on 0.2 s)\\ 
090107B & 16:20:36 & 15     &  2.3                &  X  & RT  & \\
090625B & 13:26:20 & 8      &  2.                 &  X  & RT  & \\
090702  & 10:40:37 &  6     &   0.15              &  X  & RT  & \\
090704  & 05:47:43 & 70     &  2.                 &  -  & D   & Below   threshold for automatic alert delivery.\\ 
090814B & 01:21:08 & 42     &    0.4              &  X  & RT  & \\
090817  & 00:51:23 & 30     &   2.1               &  X  & RT  & T$_{90}$ does not include bump at t$_0$+200 s\\
091015  & 23:00:17 & 100    &   0.08              &  -  & D   & Below   threshold for automatic alert delivery. \\
091111  & 15:21:59 & 100    &   0.1               &  -  & D   & Below   threshold for automatic alert delivery. \\
091202  & 23:10:04 &  25    & 0.13                &  oX & RT  & \\
091230  & 06:27:00 &  70    &   0.2               &  OX  & RT  & \\ 
100103A & 17:42:30 &  30    &    3.5              &  X  & RT  & \\
100331A & 00:30:22 & 9      &    0.5              &  -  & RT  & \\
100518A & 11:33:37 & 25     &    0.5              &  OX  & RT  & z=4 (photometric) \\
100703A & 17:43:37 & 0.06    &    2                &  -  &  D  & Short (peak flux on 0.01 s),  \\
          &     &      &                 &    &                &  Below threshold for automatic alert delivery\\
100713A & 14:35:50 &   20   &    0.4              &  X  & RT  & \\
100909A & 09:04:00 &   60   &    0.14             &  OX  &  D  & Below threshold for automatic alert delivery\\
100915B & 05:49:38 &  4     &   0.8               &  -  & RT  & \\
101112A & 22:10:20 &  6     &   1.6               &  rOX & RT  & \\
110112B & 22:24:55 &   0.3   &    5.               &  -  & D   & Short (peak flux on 0.1 s),  \\
          &     &      &                 &    &                &  Below threshold for automatic alert delivery\\
110206A & 18:08:10 &  15    &    1.6              & OX   & RT  &  \\
110708A & 04:43:30 &  50    &   0.8               & ox & RT  &  \\
110903A & 02:38:30 &   430  &  3.4                &  X  & RT  & \\
120202A & 21:40:00 &   70   &    0.2              &  -  & RT  & \\
120419A & 12:56:40 & 15     &   0.4               &  x & RT  & \\
120512A & 02:41:40 & 20     &   5.                &  -  & D   & Below threshold for automatic alert delivery \\
120711A & 02:44:38 &  135   &   27.               & OX   & RT  & \\
120821A & 13:23:35 &   12   &  0.7                &  x & RT  & \\
121102A & 02:27:02 &  25    &    1.8              &  X  & RT  & Also Swift/BAT \\
\hline
\end{tabular}
\end{footnotesize}
\caption{Continuation of Table 1.}
\label{tab1b}
\end{table}

\section{Acknowledgments}

I would like to thank J.Borkowski, D. G{\"o}tz and all the other scientists and software engineers
who contributed to the development of IBAS, as well as all the  operational staff at the ISDC, coordinated by
C.Ferrigno, for their continuous support.


\begin{thebibliography}{99}
 
\bibitem{mer03}
 {Mereghetti}, S., {G{\"o}tz}, D.,    Borkowski, J., Walter, R.,     Pedersen, H., A\&A, 411, L291, 2003

\bibitem{cou03}
 Courvoisier, T.~J.-L., et al.,  A\&A, 411, L53, 2003

\bibitem{ibis} 
Ubertini P., Lebrun F., Di Cocco G., et~al, A\&A 411, L131, 2003

\bibitem{spi} 
Vedrenne G., Roques J.-P., Sch\"{o}nfelder V., et~al, A\&A 411, L63, 2003

\bibitem{jemx}
Lund N., Budtz-Jorgensen C., Westergaard N.J., et~al., A\&A 411, L231, 2003

\bibitem{omc} 
Mas-Hesse J.M., Gim\'{e}nez A., Culhane L., et~al, A\&A 411, L261, 2003


\bibitem{via09}
 {Vianello}, G., {G{\"o}tz}, D.,   {Mereghetti}, S., A\&A, 495,  1005, 2009

\bibitem{fol08}
Foley, S., McGlynn, S., Hanlon, L., McBreen, S.,   McBreen, B.,  A\&A, 484, 143, 2008 

\bibitem{mer12}
{Mereghetti}, S., {Esposito}, P., {Tiengo}, A., {G{\"o}tz}, D., {Israel}, G.~L., {De Luca}, A., A\&A, 546, A30, 2012 

\bibitem{cra70}
{Crawford}, D.~F. and {Jauncey}, D.~L. and {Murdoch}, H.~S., ApJ, 162, 405, 1970
  

\end{thebibliography}
\end{document}